\documentclass[9pt,twocolumn,twoside]{osajnl}

\journal{ao} 

\setboolean{shortarticle}{true}

\title{Formation of gigahertz pulse train by chirped terahertz pulses interference}

\author[1]{Xinrui Liu}
\author[1,*]{Maksim Melnik}
\author[1]{Egor Oparin}
\author[1]{Maria Zhukova}
\author[1,2,3]{Joel J.~P.~C. Rodrigues}
\author[1]{Anton Tcypkin}
\author[1]{Sergei Kozlov}

\affil[1]{International Laboratory of Femtosecond Optics and Femtotechnologies, ITMO University, St. Petersburg, Russia}
\affil[2]{Federal University of Piau\'i, Teresina, PI, Brazil} 
\affil[3]{Instituto de Telecomunica\c{c}\~{o}es, Portugal} 

\affil[*]{Corresponding author: mmelnik@itmo.ru}

 \doi{\url{}}

\begin{abstract}
The development of multiplexing information transmission technology in the THz band may accelerate the arrival of the 6G era. In this paper, has demonstrated the feasibility of forming a sequence of subpulses in the temporal domain and the corresponding quasidiscrete spectrum in the THz frequency range. It is shown that despite the fact that the THz pulse has an exponential chirp, there is a "linkage relation" between spectrum and temporal structures of the THz pulse train. This fact can be used for encoding information in THz pulses for implementation in 6G communication systems in the future.

\end{abstract}

\setboolean{displaycopyright}{false}

\begin{document}

\maketitle

The maturity and commercialization of the fifth generation (5G) communications has already arrived. 6G -- a technology that is considered 100 to 1000 times \cite{andrews2014will} faster than 5G, is considered to come to our lives in the next ten years \cite{yastrebova2018future}. Fresh spectral bands as well as advanced physical layer solutions are required for future wireless communications \cite{zhang20196g}. A lot of different R$\&$D projects for the 6G network technology have been launched all over the world in the past two years \cite{zong20196g, yang20196g}. Due to the unique characteristics of the terahertz (THz) band, THz technology meets and satisfies the 6G requirements \cite{rappaport2019wireless, xing2018propagation}: ultra-wide band (0.06-10 THz for THz range) \cite{cacciapuoti2018beyond}, ultra-high data rate (up to 1 Tbps) \cite{letaief2019roadmap}, and low-latency communications (less than 0.1 ms) \cite{strinati20196g}. Now terahertz technology is considered to be a powerful supporter of the 6G network in physical layer, not only just a candidate. Terahertz massive multiple input multiple output (MIMO) antenna plays an important role in “beyond 5G” \cite{busari2019terahertz, huq2019terahertz, boulogeorgos2018terahertz}. The implementation of these “ultra-techniques” relies on ultra-fast coding and decoding methods on terahertz wave. Moreover, pulsed broadband THz technologies can also contribute to this area. For instance, quasi-discrete THz supercontinuum, obtained via interference of two THz pulses, can be used to achieve data transfer rate of 34.1 Gb/s with 31 THz spectral lines \cite{grachev2017wireless}. One of the promising techniques for data encoding is implemented by cutting out the spectral lines in quasi-discrete spectrum which correspond to the separate sub-pulses in temporal pulse train. Perspectives of this method was both experimentally \cite{10.1117/1.OE.54.5.056111, tcypkin2017spectral} and theoretically \cite{melnik2019analysis} demonstrated. However, method mentioned in these works was implemented only in NIR range. In this paper, we improved the method to work in the terahertz range by utilizing two chirped THz pulses. The corresponding relationship between the temporal and spectral pulses of sequence is demonstrated. In spite of a not linear, but an exponential chirp, there is a "linkage relation" between the emerging temporal and spectral structures - changes in spectrum will lead to similar changes in temporal domain. This technique allows to create communication network and devices which can operate at room temperature.

\begin{figure}[hb!]
\centering
\fbox{\includegraphics[width=\linewidth]{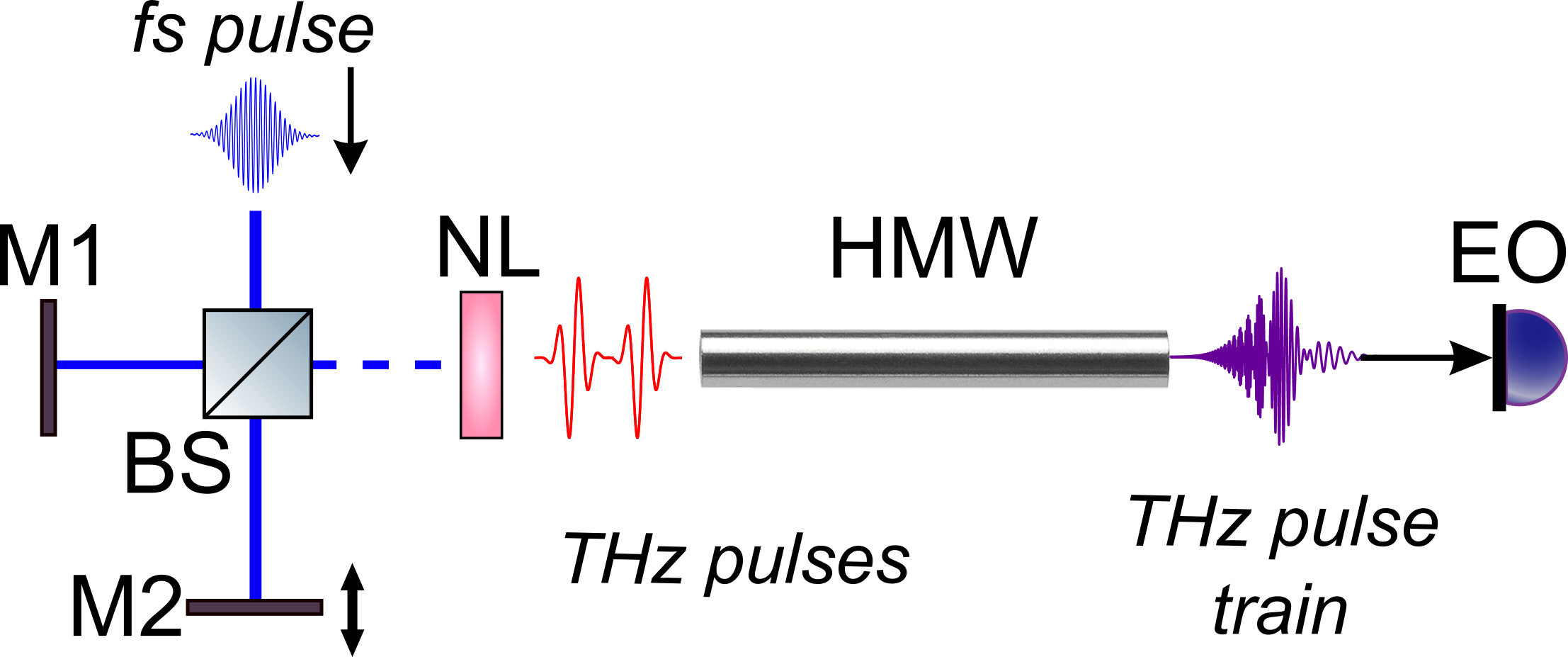}}
\caption{Experimental setup for THz pulse train generation from interference of two chirped THz pulses based on the conventional THz time-domain spectrometer. BS  -- beamsplitter, M1--M2 -- Michelson interferometer mirrors, NL -- nonlinear crystal for optics--to--THz conversion, HMW -- hollow metal waveguide, EO -- electro-optical detection system.}
\label{fig1}
\end{figure}

Figure \ref{fig1} illustrates the experimental setup for THz pulse train generation from interference of two chirped THz pulses based on the conventional THz time-domain spectrometer \cite{Grachev_2014}. In this system, the THz radiation is generated by the optical rectification of femtosecond pulses in an InAs crystal located in 2.4 T magnetic field \cite{bespalov2008methods}. The Yb-doped solid-state fs oscillator (central wavelength 1050 nm, duration 100 fs, pulse energy 70 nJ, repetition rate 70 MHz) is used as a pump. The THz radiation has estimated average power 30 $\mu$W, FWHM $\sim$2 ps. [100]-oriented CdTe crystal is used for electro-optical detection.

\begin{figure}[h!]
\centering
\fbox{\includegraphics[width=\linewidth]{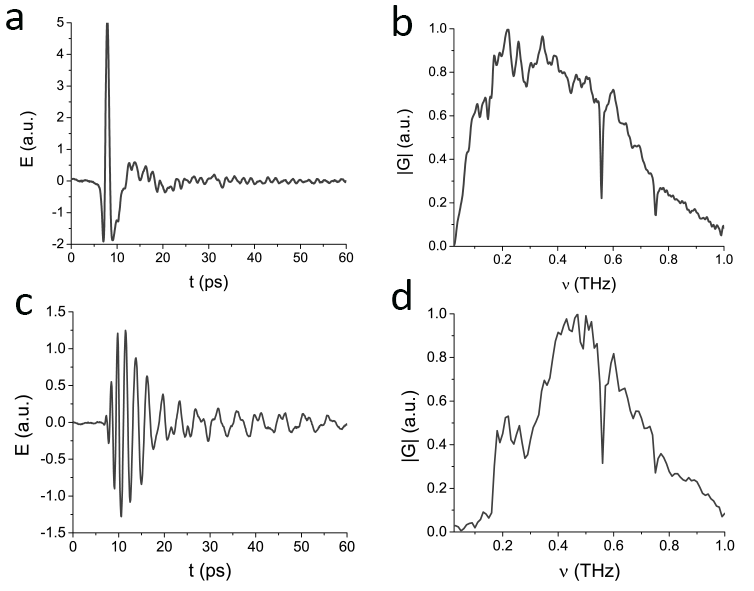}}
\caption{(a) Single THz pulse generated in InAs and (b) its spectrum. (c) The chirped pulse from hollow metal waveguide and (d) its spectrum.}
\label{fig2}
\end{figure}

It is known that two-beam interference leads to quasi-discrete spectrum spectrum \cite{grachev2017wireless}. In this work Michelson interferometer in front of a THz generator is used to create two consecutive THz pulses. One of the mirror is fixed while the other is located on the linear stage which allow to ajust the time delay between fs pulses. Then these pulses pass through the hollow metal waveguide, where two consecutive phase-modulated pulses are formed \cite{McGowan:99}. Figure \ref{fig2}(a) and (b) show example of generated single Thz pulse and its spectrum and figure \ref{fig2}(c) and (d) illustrate the chirped THz pulse obtained from hollow stainless steel metal waveguide with 23 mm length, 0.89 mm tip inner diameter, and  1.43 mm outer diameter.

As can be seen from the figure \ref{fig2}, chirping in a metal waveguide leads to an increase in the THz pulse duration from 2 ps to 7 ps, while the corresponding spectrum undergoes only minor changes. For example, the water absorption line at a frequency of 0.55 THz remains unchanged. However, at low frequencies there is a drop due to the fact that at these frequencies the signal does not propagate in the waveguide with such parameters. The 0.3 THz dip is also caused by the propagation  features of THz radiation in the waveguide \cite{McGowan:99, Gallot:00, 8281070}.

\begin{figure}[h!]
\centering
\fbox{\includegraphics[width=\linewidth]{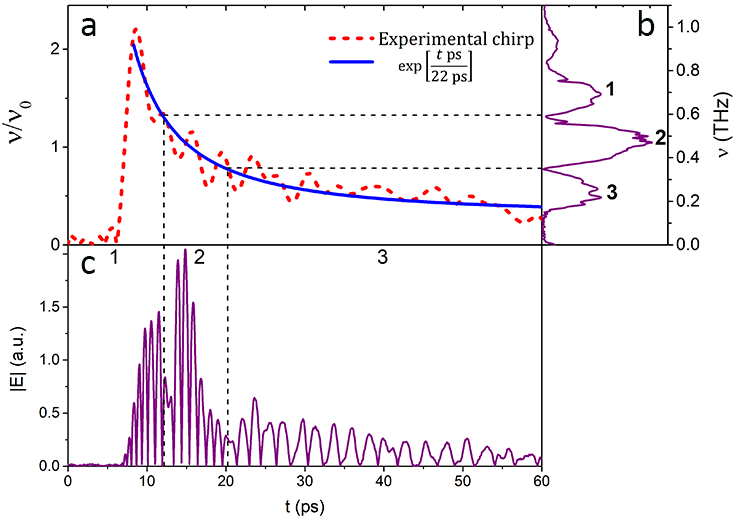}}
\caption{(a) Experimental chirp of Thz pulse and its exponential approximation, (b) quasi-discrete spectrum (c) and temporal structure of the pulse train.}
\label{fig3}
\end{figure}

The presence of the chirp makes it possible to observe not only the formation of a quasi-discrete spectrum \cite{grachev2017wireless}, but also the formation of a train of pulses during the interference \cite{tcypkin2017spectral,melnik2019analysis}. The chirp of experimentally obtained THz pulses is shown in figure \ref{fig3}(a) (red curve). As can be seen, it is well approximated by an exponential function (blue curve). Figure \ref{fig3}(b) and (c) illustrates the experimental quasi-discrete spectrum and the temporal structure of the formed pulse train.

It was previously shown \cite{tcypkin2017spectral, melnik2019analysis} that for the case of linearly chirped pulse with itself shifted by a time delay shorter than its duration interference quasi-discrete spectrum and pulse train formed have strict correspondence. This means that each subpulse in the temporal structure has its own spectral line in the quasi-discrete spectrum. However, such a correspondence was proved \cite{melnik2019analysis} only in the case of quasi-linear chirp. In this paper, the chirp is exponential, so the presence of a correspondence between the temporal and spectral structures is not obvious. Despite this, it can be seen in figure \ref{fig3} that the same number of substructures can be distinguished in the structures formed, which may indicate such a correlation. In the THz spectrum in figure \ref{fig3}(b) three discernable pulse spikes can be seen; in the temporal structure \ref{fig3}(c) these spikes correspond to three pulses of different frequencies. The time interval between the maxima of the first two pulses of a higher frequency is 10 ps, that means their repetition rate is 100 GHz.

To verify the assumptions made during the analysis of experimental data, a numerical simulation was carried out with parameters close to experimental ones. THz pulse with exponential chirp can be represented as:

\begin{equation}
E = E_0 \cdot exp(-2\frac{t^2}{\tau_0^2}) \cdot sin(2\pi\nu_0(1+exp(-\alpha t))t)
\label{eq1}
\end{equation}
where $E_0$ is the pulse amplitude, $\omega_0$ is the pulse central frequency, $\tau_0$ is the pulse duration, $\alpha$ is the inverse steepness of exponential chirp. To match the experiment, these parameters were chosen as follows: $\nu_0$ = 0.45 THz, $\tau_0$ = 7 ps and $\alpha$ = $1/22$ ps$^{-1}$.

\begin{figure}[h!]
\centering
\fbox{\includegraphics[width=\linewidth]{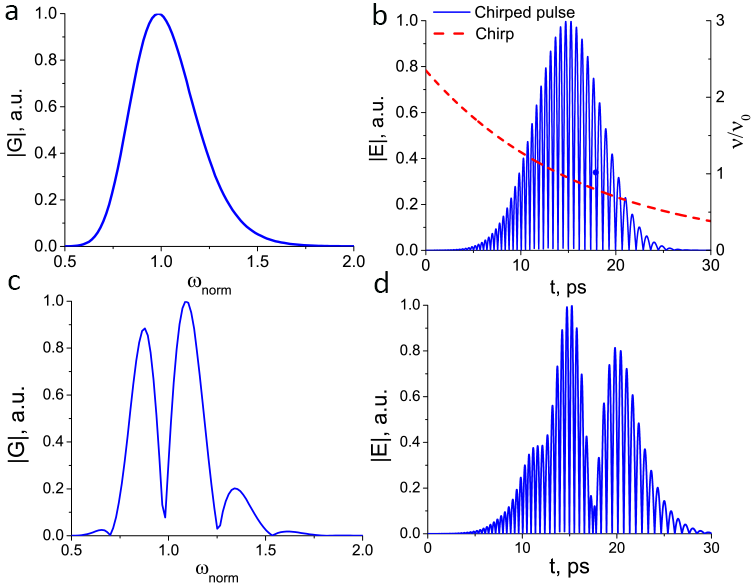}}
\caption{(a) Spectrum, (b) temporal structure and chirp of simulated chirped THz pulse. (c) quasi-discrete spectrum and (d) pulse train formed from interference of two chirped pulses.}
\label{fig4}
\end{figure}

Figure \ref{fig4}(a,b) illustrates the THz pulse with exponential chirp spectrum and temporal structure obtained from numerical simulation respectively. The interference of two such chirped pulses can be represented as \cite{tcypkin2017spectral, melnik2019analysis}:

\begin{equation}
E_{sum} = E(t)+E(t+\Delta t) 
\label{eq2}
\end{equation}
where $\Delta t$ is the time delay between pulses.

\begin{figure}[h!]
\centering
\fbox{\includegraphics[width=\linewidth]{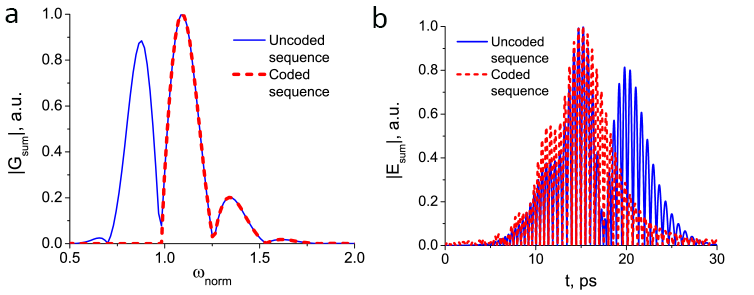}}
\caption{One of comparison example of uncoded (blue solid line) and coded (red dashed line) pulse train with cutting out one of the lines (a) in quasi-discrete spectrum and (b) corresponding pulse train changes.}
\label{fig5}
\end{figure}

Figure \ref{fig4}(c,d) represents such pulse interference with itself shifted on the time delay $\Delta t$ = 4 ps and the corresponding quasi-discrete spectrum. It can be seen that simulation results are pretty similar to the experimental one. The existing discrepancy is due to the fact that the experimental pulse has a non-Gaussian profile. To confirm the correspondence between the temporal and spectral structures formed during the interference, a numerical simulation of the cutting out one of the lines of the quasi-discrete spectrum was performed. The results are shown in Figure \ref{fig5}.

It can be seen that cutting out one of the spectral peaks leads to the vanishing of the subpulse in temporal structure. However, there is some ambiguity in temporal domain which can be explained by the presence of a residual interference term. This term can be eliminated by the proper selection of experimental parameters \cite{melnik2019analysis}. Thus, changes in spectrum will lead to similar changes in temporal domain. In other words there is "linking relation" between spectrum and temporal structures of THz pulse train. Since this special "linkage relation" between spectrum and temporal domain of chirped THz pulse train, some narrow band spectral filters can be used in the future work take an important part of terahertz information encoding and information transfer system.

In conclusion, this paper has shown the feasibility of forming a sequence of subpulses with a 100 GHz frequency in the temporal domain and the corresponding quasidiscrete spectrum in the THz frequency range. It is shown that despite the fact that the THz pulse has an exponential chirp, there is a "linkage relation" between the temporal and spectral structures. This fact can be used in the future for encoding information using such THz pulses for implementation in 6G communication systems.

\section*{Funding}

Government of the Russian Federation (08-08).

\section*{Acknowledgment}

This work was partially supported by the National Funding from the FCT - Funda\c{c}\~ao para a Ci\^encia e a Tecnologia through the UID/EEA/$50008$/$2019$ Project; by the Government of the Russian Federation, Grant $08$-$08$; and by Brazilian National Council for Research and Development (CNPq) via Grants No. $431726/2018-3$ and  $309335$/$2017$-$5$.

The authors acknowledge Dr. M. Nazarov from NRC «Kurchatov Institute» for fruitful discussions. 

\bigskip

\bibliographystyle{abbrv}
\bibliography{sample}

\end{document}